\documentclass[aps,prb,twocolumn,showpacs,superscriptaddress]{revtex4}

\usepackage{amsmath}
\usepackage{amsfonts}
\usepackage{amssymb}
\usepackage{graphicx}
\usepackage{dsfont}
\usepackage{sidecap}

\newcommand{\abs}[1]{\ensuremath{\left\vert#1\right\vert}}

\renewcommand{\Re}[1]{\ensuremath{\operatorname{Re} \left\{#1\right\} } }
\renewcommand{\Im}[1]{\ensuremath{\operatorname{Im} \left\{#1\right\} } }

\renewcommand{\vec}[1]{\boldsymbol{#1} }

\begin{document}

\title{Plasmons and screening in a monolayer of MoS$_2$}
\author{Andreas Scholz}
\email[To whom correspondence should be addressed. Electronic
address:]{andreas.scholz@physik.uni-regensburg.de}
\affiliation{Institute for Theoretical Physics,
University of Regensburg, D-93040 Regensburg, Germany}
\author{Tobias Stauber}
\affiliation{Departamento de Fisica de la Materia Condensada
and Instituto Nicolas Cabrera,
Universidad Autonoma de Madrid, E-28049 Madrid, Spain}
\author{John Schliemann}
\affiliation{Institute for Theoretical Physics,
University of Regensburg, D-93040 Regensburg, Germany}
\date{\today}

\begin{abstract}
We investigate the dynamical dielectric function of a monolayer of molybdenum disulfide
within the random phase approximation.
While in graphene damping of plasmons is caused by interband transitions,
due to the large direct band gap in monolayer MoS$_2$ collective charge excitations
enter the intraband electron hole continuum
similarly to the situation in two-dimensional electron and hole gases.
Since there is no electron-hole symmetry in MoS$_2$,
the plasmon energies in \textit{p}- and \textit{n}-doped samples clearly differ.
The breaking of spin degeneracy caused by the large intrinsic spin-orbit interaction
leads to a beating of Friedel oscillations for sufficiently large carrier concentrations, for holes as well as for electrons.
\end{abstract}


\maketitle

\section {Introduction}
\label{sect:Introduction}

Since the first isolation and detection of a monolayer of graphite,\cite{Novoselov_2004}
a system with exceptional electronic properties,
an intense search for other truly two-dimensional materials has begun.
Though graphene is widely believed to play an important role for novel electronic devices,
one of its main disadvantages is the absence of a band gap.\cite{Wallace_1947}
To overcome this problem, several proposals described ways in order to create such a gap,
e.g., by putting graphene on a certain substrate\cite{Zhou_2007} or applying a radiative field.
\cite{Calvo_2011, Oka_2009, Scholz_2013_2}
Moreover, as graphene is formed by carbon atoms, spin-orbit coupling (SOC)
is naturally small\cite{Kane_2005, Gmitra_2009} and it remains questionable whether one can
take advantage of spin-related phenomena in graphene, even though several authors described ways to enlarge
the effects of spin-orbit interactions (SOIs) in graphene significantly
by changing its environment.\cite{Neto09, Weeks_2011, Ma_2012, Varykhalov_2008, Dedkov_2008}

Another two-dimensional system that attracted a lot of attention recently
is a monolayer of molybdenum disulfide (ML-MDS),\cite{Mak_2010, Radis_2011, Zeng_2012}
a honeycomb lattice made of molybdenum and sulfur atoms instead of carbon.
The electronic properties of the monolayer differ significantly from that of bulk MoS$_2$;
e.g., while the former has a direct band gap, the latter is known to be an indirect semiconductor.
\cite{Cappelluti_2013, Kosminder_2013}
Contrary to graphene, the band gap in ML-MDS separating the valence and conduction bands 
is naturally large and due to the absence of inversion symmetry in ML-MDS
the intrinsic SOC parameter turns out to be three orders of magnitude larger
than in graphene; i.e., $\lambda\approx80$meV.

One possible application of graphene and related materials discussed in the literature
could be as a plasmonic circuit,\cite{Bonaccorso_2010, Koppens_2011,Bao_2012, Grigorenko_2013}
where density waves created by an incident light beam carry optical signals through a nanowire.
For this a better understanding of the dynamics of the collective charge excitations 
and thus of the dielectric function is indispensable.
Moreover, the dielectric function will not only be relevant for plasmonics
but also for transport and for the phonon spectra\cite{Kaasbjerg_2013} as its static limit
determines the screening behavior of the Coulomb potential.\cite{Castellanos_2013}

In recent years large effort has been made in the discussion of the dielectric function of graphene
under various conditions.
\cite{Wunsch_2006, Hwang_2007, Pyat_2009, Stauber_2010, Stauber_2010_2, Stauber_2010_3, Scholz_2011, Scholz_2012}
One of the main findings was that the behavior of plasmons in graphene
in several aspects is quite different compared to \textit{traditional} two-dimensional materials such as
III-V semiconductor quantum wells
\cite{Stern_1967, Pletyuhkov_2006, Badalyan_2010, Schliemann_2010, Ullrich_2003, Scholz_2013}
due to the relativistic nature of the charge carriers and the existence of a pseudospin degree of freedom.
Although in both systems, ML-MDS and graphene, the atoms are arranged in a honeycomb lattice,
with two inequivalent corners of the Brillouin zone denoted as valleys,
the energy spectrum in ML-MDS turns out to be quite different compared to that of graphene as
in the former electrons and holes cannot be considered as massless particles
but rather carry a finite effective mass
due to the large band gap being of the order of the hopping parameter.
Hence, we expect the dielectric function in ML-MDS to share features of
both graphene and a two-dimensional electron gas.

This work is organized as follows.
In Sec.~\ref{sect:Model}, we introduce the low-energy model Hamiltonian for ML-MDS
and summarize the formalism of the random phase approximation (RPA).
In Sec.~\ref{sect:Plasmons}, the plasmon spectra for
the \textit{n}- and \textit{p}-doped cases are opposed.
The oscillatory form of the asymptotic screened Coulomb potential
is analyzed in Sec.~\ref{sect:Screening}.
Finally, in Sec.~\ref{sect:Conclusions} we summarize the main results of this paper.

\section {The model}
\label{sect:Model}
We describe a monolayer of MoS$_2$ around the corners of the Brillouin zone
by the effective two-band model derived recently\cite{Rostami_2013, Kormanyos_2013}
for both spin ($s = \pm 1$) and valley ($\tau = \pm1$) components
(setting $\hbar = 1$ throughout this work):
\begin{align}
\hat H^{\tau s} &= \frac \Delta 2 \sigma_z + \tau s \lambda \frac{1-\sigma_z}{2} + t_0 a_0 \vec k \cdot \vec\sigma_\tau \notag \\
&+ \frac {k^2}{4m_0} \left( \alpha + \beta \sigma_z \right) 
+ t_1 a_0^2 \vec k \cdot \vec\sigma_\tau^* \sigma_x \vec k \cdot \vec\sigma_\tau^* .
\label{Hamiltonian}
\end{align}
Due to the large value of $\Delta = 1.9$eV a distinct energy gap of about $1.82$eV
separates the valence and conduction bands.
The intrinsic SOC proportional to $\lambda = 80$meV furthermore lifts
the spin degeneracy of the bands
For the other parameters we use\cite{Rostami_2013} $t_0 = 1.68$eV,
$\alpha = 0.43$, $\beta = 2.21$, $t_1 = 0.1$eV,
and $a_0 = a \cdot \cos{\theta}$,
where $a = 2.43 \AA$ is the length of the Mo-S bond and $\theta = 40.7^\circ$ the angle between the $x$-$y$ plane and the Mo-S bond.
As usual, $m_0$ denotes the free electron mass and $\vec \sigma_\tau = (\tau \sigma_x, \sigma_y)$ the vector 
of Pauli matrices acting on the pseudospin degree of freedom.

Equation (\ref{Hamiltonian}) is a generalization of Eq.~(3) in Ref.~\cite{Xiao_2012},
where the second line does not appear.
As mentioned in Ref.~\cite{Rostami_2013},
the terms quadratic in momentum are responsible for the inequality
of the electron and hole masses and for trigonal warping effects.
We should also mention that in Ref.~\cite{Cappelluti_2013}
the band structure of ML-MDS and multilayer MoS$_2$ has been
investigated in a combined \textit{ab initio} and tight-binding study within the full Brillouin zone,
where the resulting Hamiltonian turns out to be a further generalization of Eq.~(\ref{Hamiltonian}).
One of the findings of Ref.~\cite{Cappelluti_2013}
and of previous works\cite{Kosminder_2013, Peelaers_2012, Kormanyos_2013}
was that additional band extremes close to the \textit{K} point minima and maxima,
respectively, might be relevant for transport.
However, for the carrier densities used in the present paper,
$n = 10^{12}$ cm$^{-2}$ (typically for transport experiments such as in Ref.~\cite{Radisavljevic_2013})
and $5 \times 10^{13}$ cm$^{-2}$ (here both valence bands are filled in the \textit{p}-doped case),
we will neglect the influence of the higher bands as the additional extremes are expected to
be important only for densities larger than $10^{14}$ cm$^{-2}$
and thus the two-band model of Eq.~(\ref{Hamiltonian}) should give appropriate results.\cite{Peelaers_2012, Kormanyos_2013}

\begin{figure}[t]
\includegraphics[scale=0.275]{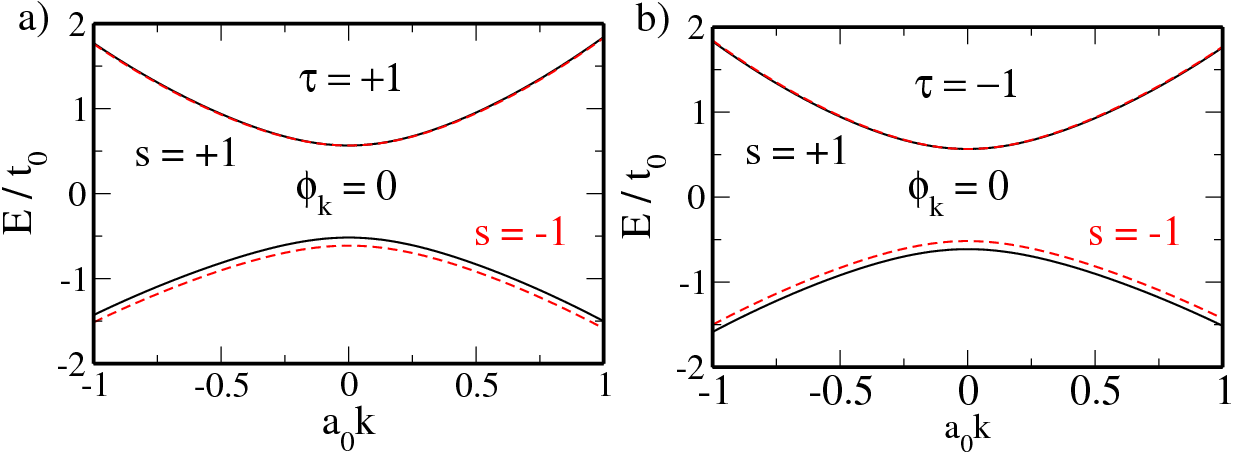}
\caption{(Color online) Energy spectrum for
(a) $\tau = +1$ and (b) $\tau = -1$
for the real spin component $s=+1$ (solid black) and $s=-1$ (dashed red).
The in-plane angle $\tan\phi_k = k_y / k_x$ was set to $\phi_k = 0^\circ$.
}
\label{FIG_Spectrum}
\end{figure}

\begin{figure}[b]
\includegraphics[scale=0.4]{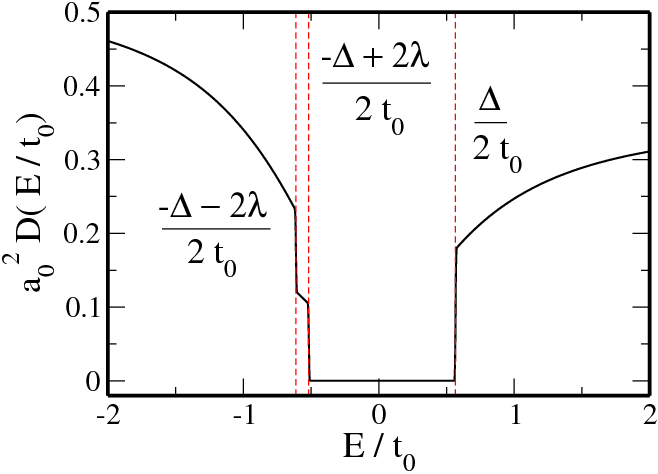}
\caption{(Color online) Density of states calculated from Eq.~(\ref{Def_DOS}).
The dashed vertical lines show the upper (lower) boundaries of the valence
(conduction) bands.
}
\label{FIG_DOS}
\end{figure}

The analytical solution of the energies obtained from Eq.~(\ref{Hamiltonian}),
\begin{align}
E^{\tau s}_\pm (\vec k) &= \frac {\alpha}{4m_0} k^2 + \frac {s \tau \lambda}2
\pm \left\{ \left( a_0^4 t_1^2 + \frac {\beta^2}{16m_0^2} \right) k^4 \right. \notag \\
& \left. + \left(\frac{\Delta-s\tau\lambda}2\right)^2
+ k^2 \left[ a_0^2t_0^2 + \frac{\beta \left(\Delta - s\tau \lambda \right)}{4m_0} \right] \right. \notag \\
& + 2\tau t_0 t_1 a_0^3 k^3 \cos{(3\phi_k)} 
\, \bigg\}^{1/2} ,
\end{align}
is shown in Fig.~\ref{FIG_Spectrum} for both valley and spin polarizations.
The trigonal warping term proportional to $t_1$ causes the spectrum to be anisotropic.
However, as $t_1 = 0.1$eV is small compared to the other energies,
this anisotropy is very weak and hence we plot only
a single in-plane angle of $\phi_k = 0^\circ$, with $\tan\phi_k = k_y / k_x$.
The valence band degeneracy is clearly broken, where for $\tau = +1$ ($\tau = -1$)
the $s = + 1$ ($s = -1$) component is energetically higher.
Because of time-reversal symmetry the corresponding shift in energy has to be opposite in the two valleys.
The conduction bands, on the other hand, remain degenerate at the \textit{K} points
but differ slightly for larger momenta due to the different curvature of the bands.
Figure \ref{FIG_DOS} displays the numerically calculated ML-MDS density of states
\begin{align}
D(E) = \int \frac {d^2k}{(2\pi)^2} \sum_{s,\tau, \sigma = \pm 1} \delta \left[ E - E^{\tau s}_\sigma(\vec k) \right] .
\label{Def_DOS}
\end{align}
Contrary to graphene the spins and valleys contribute differently to the DOS
and hence the sum over spins $s$ and valleys $\tau$ in Eq.~(\ref{Def_DOS})
cannot be replaced by a fourfold degeneracy factor.
For electron doping ($E_F > \Delta / 2$) both conduction bands are always filled,
while for hole doping either one ($-\Delta/2 - \lambda < E_F < -\Delta/2 + \lambda$)
or two ($E_F < -\Delta/2 - \lambda$) valence bands might be occupied.

In the following we want to investigate the plasmon spectrum and the screening behavior.
For this we need to calculate the dielectric function, restricting ourselves to RPA\cite{Guil_Book}
in order to account for electron-electron interactions,
given by
\begin{align}
\varepsilon(\vec q,\omega) = 1 - V(q) \chi_0(\vec q,\omega) .
\label{Def_Dielectric_Function} 
\end{align}
Here $V(q) = \frac{e^2}{2\epsilon_0 \epsilon_r q}$ is the Fourier transform
of the Coulomb potential in two dimensions,
$V(r) = \frac{e^2}{4\pi\epsilon_0 \epsilon_r r}$, $\epsilon_0$ the vacuum permittivity,
and $\epsilon_r = 5$ the background dielectric constant
(comparable to the values in Refs.~\cite{Yoon_2011, Lin_2012}).
Eq.~(\ref{Def_Dielectric_Function}) contains the free polarizability given
by a two-dimensional integral in momentum space
\begin{align}
\chi_0(\vec q, \omega) =& \sum_{s,\tau,\sigma, \sigma' = \pm 1} \int \frac {d^2k}{(2\pi)^2} 
\left|\left\langle\chi_{\sigma}^{\tau s}\left(\vec k\right) \bigg| 
\chi_{\sigma'}^{\tau s}\left(\vec k+\vec q\right) \right\rangle\right|^2 \notag \\
& \times \frac {f[E_{\sigma}^{\tau s}(\vec k)] - f[E_{\sigma'}^{\tau s}(\vec k + \vec q)]}
{\omega - E^{\tau s}_{\sigma'}\left(\vec k+\vec q\right) + E^{\tau s}_{\sigma}\left(\vec k\right) + i0}.
\label{DEF_suscept}
\end{align}
$\left\vert \chi_{\sigma}^{\tau s}\left(\vec k\right) \right\rangle$
and $E^{\tau s}_{\sigma}\left(\vec k\right)$ are the eigenstates and
energies for a given valley ($\tau$), spin ($s$), and pseudospin ($\sigma$).
Notice that only one sum over $s$ and $\tau$,
respectively, appears in Eq.~(\ref{DEF_suscept}) as spin or valley changing transitions are forbidden.
In the following we assume zero temperature. The Fermi function $f[E]$ then reduces to a simple step function.

\begin{figure}[t]
\includegraphics[scale=0.55]{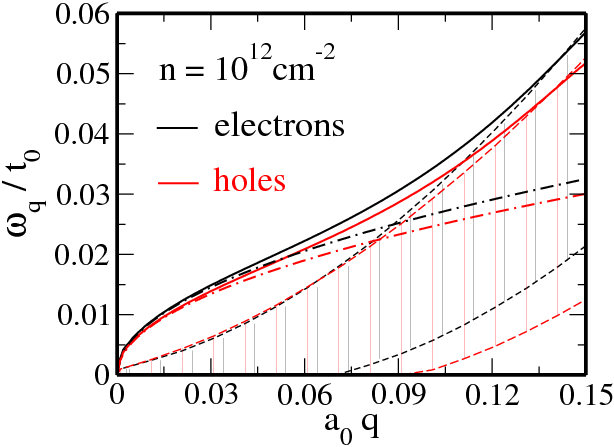}
\caption{(Color online) The solid lines show the plasmon spectrum for an electron (black)
and hole (red) concentration of $n = 10^{12}$ cm$^{-2}$.
The dashed lines show the boundaries of the EHC.
The dotted-dashed lines are the long-wavelength results of Eq.~(\ref{long_wavelength_plasmon}).
The in-plane angle was set to $\phi_q = 0^\circ$.
}
\label{FIG_Plasmons_low_doping}
\end{figure}

For the special case of $\alpha = \beta = t_1 = 0$
(corresponding to the model of Ref.~\cite{Xiao_2012}),
the above expression (\ref{DEF_suscept})
equals that of gapped graphene,\cite{Pyat_2009, Scholz_2011}
where each contribution with valley $\tau$ and spin $s$ has to be described 
with an effective mass term of $\tilde\Delta^{\tau s} = \Delta / 2 - s\tau \lambda / 2$
and a shifted Fermi energy of $\tilde E_F^{\tau s} = E_F - s\tau \lambda / 2$.
In the following, however, we do not neglect the terms quadratic in momentum
but rather solve $\varepsilon(\vec q, \omega)$ within the extended model of Eq.~(\ref{Hamiltonian}).
This is done numerically by first calculating the imaginary part of the polarizability
using the Dirac identity $\Im{1 / (x \pm i0)} = \mp \pi \delta(x)$.
Afterwards the result is integrated with the help of the Kramers-Kronig relation
\begin{align}
\Re{\chi_0(\vec q, \omega)} = 
\frac 2\pi \mathcal P \int_0^\infty d\omega' \frac {\omega' \Im{\chi_0(\vec q,\omega')}}{\omega'^2-\omega^2}
\label{KKR}
\end{align}
to obtain the real part.

\section {Collective charge excitations}
\label{sect:Plasmons}

In the case in which the dielectric function in Eq.~(\ref{Def_Dielectric_Function}) vanishes,
\begin{align}
\varepsilon (\vec q, \omega_q) = 0 , \label{Def_Plasmons}
\end{align}
the system exhibits characteristic density waves known as plasmons.
If the quasiparticle energy $\omega_q$ is large compared to the damping rate,
the complex valued Eq.~(\ref{Def_Plasmons}) can further be substituted
by the approximate equation\cite{Guil_Book}
\begin{align}
\Re{\varepsilon (\vec q, \omega_q) } = 0 . \label{Def_weakly_damped_Plasmons} 
\end{align}
Only if the solution $\omega_q$ additionally corresponds to a resonance
in the energy loss function,
$-\Im{1 / \varepsilon(\vec q, \omega_q + i0)}$,
a quantity which is available in scattering experiments,
one can speak of a long-lived coherent mode.

\begin{figure}[b]
\includegraphics[scale=0.5]{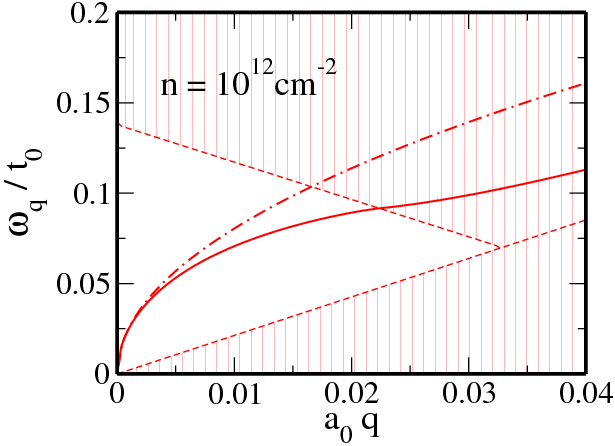}
\caption{(Color online) Plasmon dispersion (solid line) and
boundaries of the EHC (dashed) for graphene with $n=10^{12}$ cm$^{-2}$.
The dotted-dashed line shows the long-wavelength result 
$\omega_q^{0,g} = \sqrt{e^2 E_F q / 2\pi\epsilon_0}$.
}
\label{FIG_Plasmons_Graphene}
\end{figure}

In the long-wavelength limit the analytical expression\cite{Sensarma_2010}
for the plasmon dispersion reads (neglecting the trigonal warping term for the moment)
\begin{align}
\omega_{q, \pm}^0 = \sqrt{\frac{e^2}{8\pi \epsilon_0\epsilon_r}
\sum_{\tau,s=\pm1} k_F^{\tau\cdot s} \abs{\frac{\partial E_\pm^{\tau s}}{\partial k}}_{k = k_F^{\tau\cdot s}} } \, \sqrt{q} ,
\label{long_wavelength_plasmon}
\end{align}
with the universal $\sqrt q$ dependence of two-dimensional materials.
Here the upper (lower) sign stands for the \textit{n}-doped (\textit{p}-doped) case.
The Fermi wave vector in Eq.~(\ref{long_wavelength_plasmon}) is given by
\begin{align}
k_F^{\pm} &= \frac{\sqrt8 m_0 a_0 t_0}{\sqrt{\beta^2-\alpha^2}} \text{Re} \Bigg[
\left\{ -1 - \frac{2\alpha E_F + \beta\Delta \mp \left(\alpha + \beta\right) \lambda }{4m_0a_0^2t_0^2} \right. \notag \\
& \left. + \left[\left(\beta^2-\alpha^2\right)\frac{\left(2E_F - \Delta\right)
\left(2E_F + \Delta \mp 2\lambda\right)}{16m_0^2a_0^4t_0^4} \right. \right. \notag \\
& \left. \left. + \left( 1 + \frac{2\alpha E_F + \beta\Delta \mp \left(\alpha+\beta\right)\lambda}{4m_0a_0^2t_0^2} \right)^2
\right]^{1/2}
\right\}^{1/2} \Bigg] .
\end{align}

\begin{figure}[b]
\includegraphics[scale=0.55]{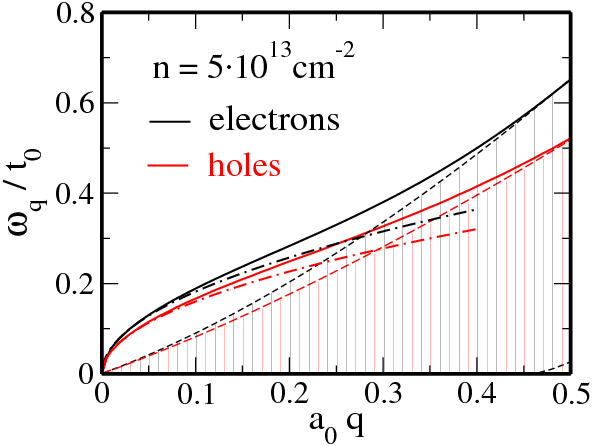}
\caption{(Color online) The solid lines show the plasmon spectrum for an electron (black)
and hole (red) concentration of $n = 5 \times 10^{13}$ cm$^{-2}$.
The dashed lines show the boundaries of the EHC.
The dotted-dashed lines are the long-wavelength results of Eq.~(\ref{long_wavelength_plasmon}).
The in-plane angle was set to $\phi_q = 0^\circ$.
}
\label{FIG_Plasmons_high_doping}
\end{figure}

Due to the electron-hole symmetry in graphene,
plasmons in \textit{n}- and \textit{p}-doped samples at a given
carrier concentration show the same dynamics.
This is obviously no longer true in ML-MDS
as the structure of the valence bands is quite different
compared to the conduction bands; see Fig.~\ref{FIG_DOS}.

In Fig.~\ref{FIG_Plasmons_low_doping}
the plasmon dispersion and the intraband part of the electron-hole continuum (EHC) are shown
at a given carrier concentration of
$n = \sum_{\nu = \pm1} (k_F^{\nu})^2 / 2 \pi = 10^{12}$ cm$^{-2}$
for electron (black) and hole (red) doping.
The in-plane angle orientation was set to $\phi_q = 0^\circ$,
where $\tan\phi_q = q_y / q_x$.
The dotted-dashed lines show the long-wavelength result of Eq.~(\ref{long_wavelength_plasmon}),
which turns out to be in good agreement with the numerical solution for $a_0 q \lesssim 0.05$.
The plasmon dispersions and the EHC for \textit{n} and \textit{p} doping clearly differ,
where $\omega_q$ is energetically higher in the former.

Due to the large value of the band gap $\Delta$,
the interband part of the EHC in ML-MDS is energetically very high
and, subsequently, the plasmon dispersion enters the intraband EHC.
This is quite different compared to graphene
where due to the singularity of the free polarizability at $\omega = v_F q$
(with $v_F = 10^6$m/s being the Fermi velocity in graphene)
damping can only be caused by interband transitions.\cite{Wunsch_2006}
Comparing, e.g., Fig.~\ref{FIG_Plasmons_low_doping} with the corresponding result
obtained for suspended graphene (Fig.~\ref{FIG_Plasmons_Graphene}),
we can immediately see that the mode in graphene becomes damped at much smaller wave
vectors $a_0 q \approx 0.02$ compared to ML-MDS where damping appears not before $a_0q \approx 0.15$.
Moreover, the energy loss function of graphene for such large momenta does not exhibit a 
resonant peak and thus the plasmon is already overdamped.
However, the plasmon energies in graphene are clearly larger compared to ML-MDS,
e.g., $\omega^{gr}_q/t_0 \approx 0.09$, while $\omega^{MoS_2}_q/t_0 \approx 0.01$
at $a_0 q = 0.02$.
It is interesting to note that while the long-wavelength result in graphene,\cite{Wunsch_2006}
$\omega_q^{0,g} = \sqrt{e^2 E_F q / 2\pi \epsilon_0}$
(see dashed line in Fig.~\ref{FIG_Plasmons_Graphene}),
overestimates the exact solution,
the approximate result of Eq.~(\ref{long_wavelength_plasmon})
is energetically below the numerical value.

The difference in the plasmon energies for \textit{n} and \textit{p} doping
becomes enhanced for larger $n = 5\times 10^{13}$ cm$^{-2}$
as the difference in the electron and hole masses becomes more important;
see Fig.~\ref{FIG_Plasmons_high_doping}.
\begin{figure}[t]
\includegraphics[scale=0.45]{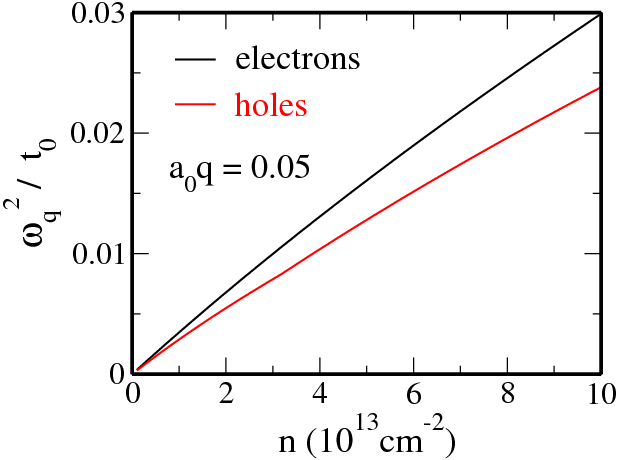}
\caption{(Color online) Density dependence of the plasmon energy
$\omega_q^2$ for electron (black) and hole (red) doping.
In both cases the spectrum clearly scales as $\omega_q \propto \sqrt{n}$.
Parameters: $a_0q = 0.05$, $\phi_q = 0^\circ$.
}
\label{FIG_Plasmons_density_dependence}
\end{figure}
A detailed analysis of the dependence of the plasmon energies $\omega_q^2$
on the carrier concentrations obtained for fixed $a_0 q = 0.05$ and $\phi_q = 0^\circ$
is shown in Fig.~{\ref{FIG_Plasmons_density_dependence},
clearly indicating that the asymmetry in the plasmon spectrum increases for larger densities.

From Fig.~{\ref{FIG_Plasmons_density_dependence} one can furthermore see that
the plasmon energy in ML-MDS is of the form $\omega_q \propto n^{1/2}$
as in a two-dimensional electron gas.
This can be understood from the long-wavelength behavior of the plasmon
frequency [neglecting for simplicity terms quadratic in momentum in Eq.~(\ref{Hamiltonian})],
\begin{align}
\omega_q^0 = \sqrt{\frac {e^2q}{2\pi\epsilon_0\epsilon_r}} \,
\sqrt{ \frac {\left(2E_F -\Delta\right) \left[E_F \left(\Delta + 2E_F \right) - \lambda^2\right] } 
{ 4E_F^2 - \lambda^2} },
\label{long_wavelength_plasmon_simplified}
\end{align}
as for realistic concentrations, e.g., $n = 10^{13}$ cm$^{-2}$,
the ratio $\Delta / 2 E_F \approx 0.97$ is close to unity and thus
we can approximate Eq.~(\ref{long_wavelength_plasmon_simplified}) by ($\lambda \ll \Delta, E_F$)
\begin{align*}
\omega_q^0 &\approx \sqrt{\frac {e^2q}{2\pi\epsilon_0\epsilon_r}} \,
\sqrt{ E_F \left[ 1 - \frac{\Delta^2}{\Delta^2 + 4\pi t_0^2 a_0^2 n} \right] } \\
&\approx \sqrt{ \frac {2e^2q\mu t_0^2 a_0^2}{\epsilon_0\epsilon_r\Delta^2} } \sqrt n .
\end{align*}
Hence ML-MDS can be considered as a kind of a non relativistic limit of gapped graphene.\cite{Scholz_2011}
In contrast, due to the ultrarelativistic nature of the charge carriers in graphene
the density dependence of the plasmon frequency $\omega_q^{0,g} = \sqrt{e^2 E_F q / 2\pi \epsilon_0}$
scales as $\omega \propto n^{1/4}$, where $n = E_F^2 / \pi v_F^2$.\cite{Hwang_2007}

Let us finally comment on the importance of the terms in Eq.~(\ref{Hamiltonian})
that are quadratic in momentum.
In Ref.~\cite{Rostami_2013, Kormanyos_2013} it was pointed that
these terms are necessary to properly describe 
the result of previous \textit{ab initio} calculations.\cite{Peelaers_2012}
Due to the smallness of the trigonal warping contribution $t_1$,
the plasmon spectrum turns out to be virtually isotropic
and the angle dependence of $\omega_q$ is negligible.
Comparing, e.g., the $\phi_q = 0^\circ$ and $\phi_q = 60^\circ$
result at a given momentum $a_0 q = 0.2$, we notice only a very small relative
difference of a few percent even for large concentrations of
$n=5\times 10^{13}$ cm$^{-2}$.
However, our calculations also show that the other contributions proportional to $\alpha$ and $\beta$,
which are responsible for the different electron and hole masses,
\cite{Rostami_2013, Kormanyos_2013, Peelaers_2012}
cannot be neglected.
Although the qualitative behavior of the plasmon dispersion is captured
by the simplified model, the energies $\omega_q$ obtained from the extended model 
turn out to be clearly enlarged even for small momenta.

\begin{figure}[b]
\includegraphics[scale=0.31]{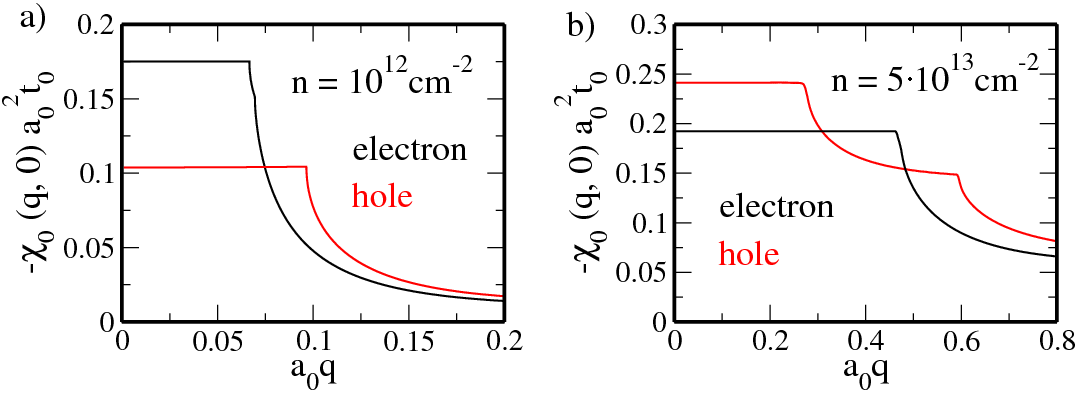}
\caption{(Color online) Static polarizability for (a) $n = 10^{12}$ cm$^{-2}$
and (b) $n = 5 \times 10^{13}$ cm$^{-2}$ for electron (black) and hole (red) doping.
}
\label{FIG_static_pol}
\end{figure}

\section {Screening of impurities}
\label{sect:Screening}

Assuming the dielectric function to be isotropic,
i.e., neglecting the trigonal warping term,
the RPA improved Coulomb potential can be obtained from
\begin{align}
\Phi(r) = \frac Q{\epsilon_0} \int_0^\infty dq \; \frac{J_0(qr)}{\varepsilon(q,0)} . \label{screened_pot}
\end{align}
Here $J_0(x)$ is the Bessel function of the first kind and $Q$ the charge of the impurity.
From the Lighthill theorem\cite{Lighthill} we know that the asymptotic behavior of $\Phi(r)$
is determined by the nonanalytical points of the dielectric function.
Right at $q = 2k_F^{\pm}$ cusps will appear in the static polarizability
indicating such singular points.

While due to the absence of backscattering on the Fermi surface
in doped graphene only the second derivative of the static dielectric function diverges
at $q = 2k_F$,\cite{Ando_2006, Wunsch_2006}
already the first derivative does in an electron gas.\cite{Stern_1967}
As a result the power-law dependence in graphene, $\Phi(r) \propto 1/r^3$,
is quite different compared to $\Phi(r) \propto 1/r^2$ in a 2DEG.
Nevertheless, in both cases the screened Coulomb potential exhibits characteristic sinusoidal
Friedel oscillations due to the existence of a sharp Fermi surface.

\begin{figure}[t]
\includegraphics[scale=0.3]{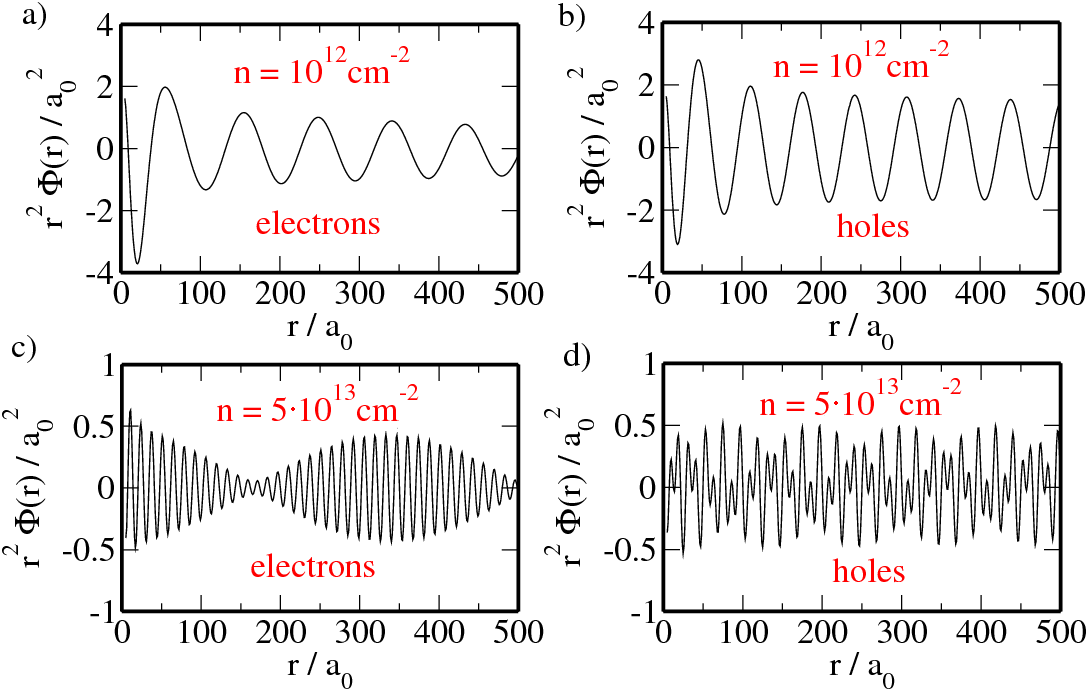}
\caption{(Color online) Numerically calculated screened potential (in units of $Qa_0 / \epsilon_0$)
for electron [(a) and (c)] and hole doping [(b) and (d)]
for two different carrier concentrations
$n = 10^{12}$ cm$^{-2}$ and $n = 5 \times 10^{13}$ cm$^{-2}$, respectively.
}
\label{FIG_screened_potential}
\end{figure}

In Fig.~\ref{FIG_static_pol} we show the static polarizability of ML-MDS for two
different concentrations $n = 10^{12}$ cm$^{-2}$ and $n = 5 \times 10^{13}$ cm$^{-2}$, respectively.
While in the former case $-\chi_0^{e}(q\rightarrow0, 0) > -\chi_0^{h} (q\rightarrow0, 0)$,
the opposite is true in the latter which can be understood from the DOS in Fig.~\ref{FIG_DOS}.

For hole densities of $n = 10^{12}$ cm$^{-2}$ only one valence band is occupied.
Hence only one Fermi wave vector is finite and
the static polarizability is singular at $q = 2k_F^+$;
see red line in Fig.~\ref{FIG_static_pol}(a).
In the other case of electron doping both conduction bands are filled
and the Fermi contour consists of two concentric circles with different radii,
where the relative difference between $k_F^+$ and $k_F^-$
(of about $5\%$) is only small.
As as result the screened potential in Fig.~\ref{FIG_screened_potential}
behaves as $\Phi(r) \propto \sin{(2k_F^+r)} / r^2$ for hole doping,
while for the electronic case $\Phi(r)$ deviates slightly from this behavior
due to an additional contribution proportional to $\sin{(2k_F^-r)} / r^2$.

The case of $n = 5 \times 10^{13}$ cm$^{-2}$ is more interesting as also in the \textit{p}-doped case
both valence bands are occupied and the corresponding wave vectors $k_F^{+}$
and $k_F^{-}$ differ significantly due to the large value of the SOC parameter;
see red line in Fig.~\ref{FIG_static_pol}(b).
The numerically calculated potential $\Phi(r)$, as shown in Fig.~\ref{FIG_screened_potential}(d),
clearly shows a superposition of two
oscillatory contributions, whose periods are given by $1/2k_F^+$ and $1/2k_F^-$, respectively.
Such a beating behavior also appears in monolayer graphene if Rashba SOIs are taken into account.\cite{Scholz_2012}
However, the important difference is that the intrinsic SOC parameter in ML-MDS of 
about $80$meV is naturally large compared to $\lambda_R = 10\mu$eV (for $1$V/nm)\cite{Gmitra_2009} in graphene
and does not need to be enlarged artificially in order to see noticeable effects.

\section{Conclusions}
\label{sect:Conclusions}

We have investigated the dynamical dielectric function in a monolayer of molybdenum disulfide.
As we have demonstrated, plasmons in ML-MDS behave similarly to those in two-dimensional electron gases.
The density dependence of the plasmon energies was shown to be of the form $\omega_q \propto n^{1/2}$,
while $\omega_q \propto n^{1/4}$ in graphene.
Moreover, damping of plasmons at large momenta is caused by the intraband transitions
and not by interband processes as in graphene.
This leads to the existence of a resonance in the energy loss function in ML-MDS for momenta where
the mode in graphene is already damped out.
Furthermore, due to the pronounced electron-hole asymmetry in ML-MDS
a distinct difference in the plasmon dispersions of \textit{n}- and \textit{p}-doped samples is predicted.
This difference was shown to increase for larger carrier concentrations.

Based on the form of the static polarizability,
we expect the screened Coulomb potential to show a beating of Friedel
oscillations for sufficiently large carrier concentrations
due to the different curvature of the conduction and valence bands with different
spin orientations.
The numerical inspection of $\Phi(r)$ confirms the above prediction,
where the period of this beating turns out to be roughly two orders of magnitude
larger than the lattice constant.

Finally, we want to point out that our results might not only be relevant for
ML-MDS but also for other group-VI dichalcogenides.
In Ref.~\cite{Xiao_2012}, for example, it has been reported that
the intrinsic SOC parameter could further be increased up to $215$meV
if the molybdenum atoms are substituted by tungsten,
which in turn would enhance the effects predicted in this work.

\acknowledgments
This work was supported by Deutsche Forschungsgemeinschaft via
Grant No.~GRK 1570,
by FCT under Grants No.~PTDC/FIS/101434/2008
and No.~PTDC/FIS/113199/2009,
and MIC under Grant No.~FIS2010-21883-C02-02.


\end{document}